\documentclass[prl, aps, twocolumn, superscriptaddress, nofootinbib, tightenlines, nobibnotes]{revtex4-1}

\usepackage{bigints}
\usepackage{amssymb}
\usepackage{graphicx}
\usepackage{dcolumn}
\usepackage{bm}
\usepackage{multirow}
\usepackage{hyperref}
\usepackage{comment}
\usepackage{xcolor}
\usepackage[caption=false]{subfig}

\definecolor{lapislazuli}{rgb}{0.15, 0.38, 0.61}
\definecolor{YKblue}{rgb}{0.0, 0.18, 0.65}
\definecolor{carmine}{rgb}{0.81, 0.09, 0.03}
\definecolor{lavender}{rgb}{0.84, 0.49, 0.87}

\hypersetup{
colorlinks=true,
linkcolor=lapislazuli,
linktoc=page,
citecolor=lapislazuli,
urlcolor=lapislazuli
}

\usepackage{amssymb}
\usepackage{mathtools}
\usepackage{multibib}

\everymath{\displaystyle} 
\newcommand{\pr}[1]{\ensuremath{\left[#1\right]}} 
\newcommand{\pc}[1]{\ensuremath{\left(#1\right)}} 
\newcommand{\bra}[1]{\ensuremath{\left\langle#1\right\vert}} 
\newcommand{\ket}[1]{\ensuremath{\left\vert#1\right\rangle}} 

\DeclareMathOperator{\Tr}{Tr} 

\begin{document}

\title{Sideband ground-state cooling of graphene with Rydberg atoms via vacuum forces}

\author{M. Miskeen Khan}
\affiliation{Instituto Superior T\'ecnico, Universidade de Lisboa, Portugal}
\affiliation{Instituto de Plasmas e Fus\~ao Nuclear, Instituto Superior T\'ecnico, Universidade de Lisboa, Portugal}
\email{miskeen.khan@tecnico.ulisboa.pt}

\author{S. Ribeiro}
\affiliation{Joint Quantum Centre (JQC) Durham-Newcastle, Department of Physics, Durham University, United Kingdom}

\author{J. T. Mendon\c{c}a}
\affiliation{Instituto Superior T\'ecnico, Universidade de Lisboa, Portugal}
\affiliation{Instituto de Plasmas e Fus\~ao Nuclear, Instituto Superior T\'ecnico, Universidade de Lisboa, Portugal}

\author{H. Ter\c{c}as}
\affiliation{Instituto Superior T\'ecnico, Universidade de Lisboa, Portugal}
\affiliation{Instituto de Plasmas e Fus\~ao Nuclear, Instituto Superior T\'ecnico, Universidade de Lisboa, Portugal}
\email{hugo.tercas@tecnico.ulisboa.pt}

\begin{abstract}

We present a scheme leading to ground-state cooling of the fundamental out-of-plane (flexural) mode of a suspended graphene sheet. Our proposal exploits the coupling between a driven Rydberg atom and the graphene resonator, which is enabled by vacuum forces. Thanks to the large atomic polarizability of the Rydberg states, the Casimir-Polder force is several orders of magnitude larger than the corresponding force achieved for atoms in the ground state. By playing with the distance between the atom and the graphene membrane, we show that resolved sideband cooling is possible, bringing the occupation number of the fundamental flexural mode down to its quantum limit. Our findings are expected to motivate physical applications of graphene at extremely low temperatures. 
\end{abstract}
\maketitle


\textit{Introduction.$-$} The interest around atomically thin mechanical resonators, such as graphene and other two-dimensional (2D) materials \cite{Bunch490,Steele2009}, has grown in the last years thanks to their low masses and large stiffnesses, leading to large oscillation frequencies and high-quality factors \cite{chen2009performance,doi:10.1021/acs.nanolett.7b01845, moser2014nanotube, doi:10.1002/andp.201400153}. Those features put graphene in the run for competitive nanomechanical solutions, with a variety of important applications in quantum technology, such as storing and processing quantum information by invoking coherent coupling \cite{luo2018strong}, high-precision quantum sensing \cite{chen2009performance, Singh2014, Muschik2014, deBonis2018}, and quantum interferometry \cite{PhysRevX.8.021052}. From the condensed matter physics perspective, the ability to control the out-of-plane vibration of a 2D material (the so-called flexural phonon) is also very challenging, as it allows for tunable control of pseudo-magnetic fields \cite{PhysRevB.77.075422}, electron-phonon coupling \cite{benyamini2014real} and selective band-gap engineering \cite{Neto_2011}. However, the quantum control of the flexural modes occurs in the near-zero thermal noise limit, where the mechanical motion is mostly due to quantum fluctuations \cite{LaHaye2004}. 
As such, ground-state cooling of graphene nanoresonators still preludes the applications mentioned above, and therefore an efficient, active cooling protocol is urgent. Since laser (or radiation pressure) and photo-thermal cooling schemes are extremely ineffective in graphene - as a result of its broad-band absorption spectrum \cite{doi:10.1021/nl302036x, PhysRevLett.108.047401, Yan2012, PhysRevLett.113.027404, DeAlba2016} - a possible solution could be exploiting the electromagnetic vacuum fluctuations (EVF), emerging as a powerful resource to couple graphene resonators with a quantum emitter. Such hybrid setups offer all-optical quantum control of the membrane motion, enabling force sensing \cite{PhysRevLett.112.223601} and mechanical squeezing \cite{PhysRevA.96.063819}. More recently, EVF has also been used for ground-state cooling of an h-BN monolayer based on electromagnetically induced transparency (EIT) \cite{PhysRevLett.122.023602}. Additionally, dispersion forces have a measurable effect on the levels and lifetimes of Rydberg atoms when brought close to surfaces \cite{PhysRevA.83.032902, PhysRevLett.108.063004}. However, solutions combining the features of both vacuum forces and high-lying Rydberg states in graphene electro-mechanical setups are still at their infancy \cite{PhysRevA.88.052521}, and the possibilities offered by Rydberg states are far from being exhausted.
\begin{figure}[!t]
\includegraphics[width=\linewidth]{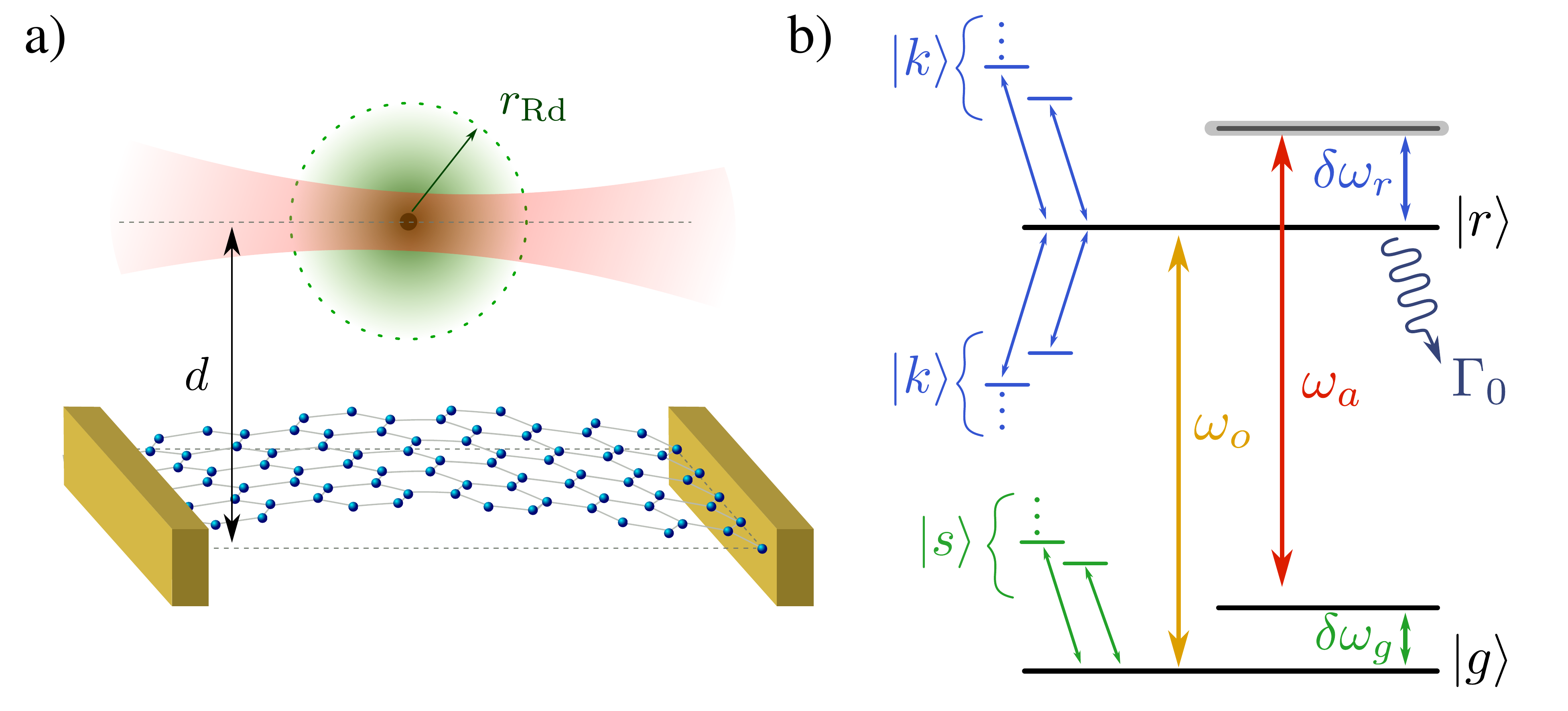}
\caption{(Color online) a) Schematic representation of the experimental setup. A Rydberg atom is placed at a distance $d$ from a suspended graphene membrane. The out-of-plane (flexural) motion results in a modulation of the Rydberg transition frequency $\omega_{a}$. b) (not in scale) The states $\vert s\rangle$ contributing to the calculation of ground-state energy shift $\delta\omega_{g}$, lying above the energy level. For Rydberg states, the energy shift $\delta\omega_{r}$ accounts for contributions from both the above- and below-lying transitions $\vert k\rangle$. In addition, the graphene resonator induces a modification in the free-space Rydberg decay rate $\Gamma_0$.} \label{cfig:fig1}
\end{figure}

In this Letter, we describe an experimentally feasible scheme to achieve ground-state cooling of a graphene resonator by coupling it to a Rydberg atom via vacuum forces. We exploit the large atomic polarizability of the Rydberg states, and the fact that the back-action of the Casimir-Polder (CP) force depends on the atom-surface distance, to construct a resolved sideband cooling protocol bringing the out-of-plane (flexural) mode down to its quantum ground state. Our findings pave the way for both quantum technological and condensed matter investigations for which ground-state cooling of flexural modes in suspended graphene (and eventually other 2D materials) is required.


\textit{Modification of the atomic transitions due to vacuum fluctuations.$-$} Our setup is depicted in Fig.~\ref{cfig:fig1}~a), where we consider an atom placed close to a graphene membrane at a distance $d$ along the $z$-direction. The interaction potential is calculated assuming the graphene sheet to be infinitely extended, thereby neglecting possible finite-size effects. Within the formalism of macroscopic quantum electrodynamics, we can write the energy shift in the ground-state $\delta\omega_{g}$, and total shift in the Rydberg state $\delta\omega_{r}$ as the sum of the non-resonant $\delta\omega_{r}^{\rm NR}$ and the resonant $\delta\omega_{r}^{\rm R}$ parts \cite{buhmann2013dispersion, scheel2008macroscopic}
\begin{align}
\delta\omega_{g} &= \frac{\hbar\mu_{0} }{2\pi} \int_{0}^{\infty} d \xi \xi^{2}{\alpha}_{g} \pc{i \xi}  \Tr \pr{\mathbf{G} \pc{\mathbf{r}, \mathbf{r}, i \xi}},
\label{eq:ucpg}\\
\delta\omega_{r}^{\rm NR}&= \frac{\hbar\mu_{0}}{2\pi}\int_{0}^{\infty} d \xi \xi^{2}{\alpha}_{r} \pc{i \xi}  \Tr \pr{\mathbf{G} \pc{\mathbf{r},\mathbf{r},i\xi}},
\label{eq:ucprNR}\\
\delta\omega_{r}^{\rm R}&=- \frac{ \mu_{0} }{3} \sum_{\ell<r } \omega^2_{r\ell}|\mathbf{d}_{r\ell}|^{2}\mathrm{Tr} \pr{\text{Re} \pr{\mathbf{G} \pc{\mathbf{r}, \mathbf{r}, \omega_{r\ell}}}}.
\label{eq:ucprR}
\end{align}
The free-space spontaneous decay rate of the Rydberg state $\Gamma_{0}$ is also modified due to the presence of the graphene sheet as 
\begin{align}
\Gamma=  \Gamma_{0}+\frac{2\mu_0}{3 \hbar}\sum_{\ell<r } \omega^2_{r\ell}|\mathbf{d}_{r\ell}|^{2}\mathrm{Tr} \pr{\text{Im} \pr{\mathbf{G} \pc{\mathbf{r}, \mathbf{r}, \omega_{r\ell}}}},
\label{eq:modificationgamma}
\end{align} 
where $\mathbf{G} \pc{\mathbf{r}, \mathbf{r}, \omega}$ is the vacuum field Green tensor accounting for the relevant electromagnetic properties of graphene (refer to the Supplementary Material located at \cite{supp} for further details). In particular, we take the values $E_{\rm F}\simeq 0.8~\text{eV}$  and $\gamma_{g}=10^{12} ~{\rm s^{-1}} $ for the graphene Fermi energy and relaxation rate, respectively \cite{chen2011controlling, PhysRevLett.105.256805, doi:10.1021/nl201771h,tielrooij2015electrical}. The properties of the atom are introduced in this formalism via the atomic polarizability function ${\alpha}_{x}\;(\omega) $, with $x=g \;(r)$ labeling the ground (Rydberg) state, and is given by
\begin{align}
\alpha_{x} (\omega) = \lim_{\epsilon \to 0} \frac{2}{3\hbar}\sum_{\ell} \frac{\omega_{\ell x} |\mathbf{d}_{x\ell}|^{2} }{\omega^2_{\ell x} - \omega^2 -i \omega \epsilon}  .
\label{eq:atompolaziability}
\end{align}
Here, the sum is extended to all possible atomic transitions, as depicted in Fig.~\ref{cfig:fig1}~b). 
\begin{figure}[t!]
\includegraphics[width=1\linewidth]{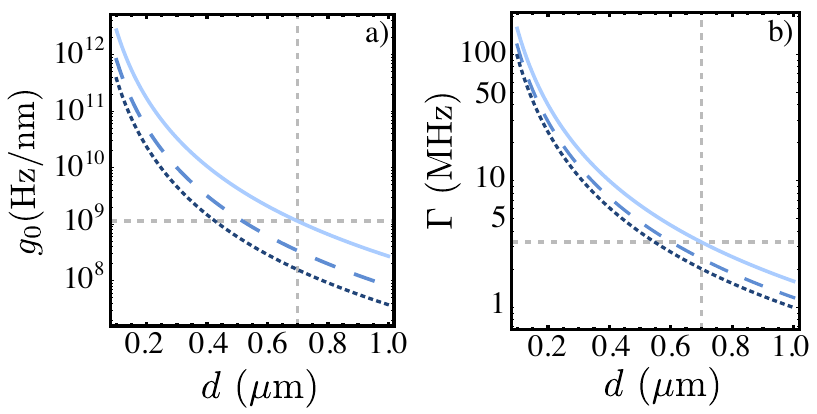}
\caption{(Color online) a) Casimir-Polder vacuum coupling $g_0$ and b) spontaneous decay rate of the Rydberg states $\ket{80 S_{1/2}}$ (solid line), $\ket{60 S_{1/2}}$ (dashed line) and $\ket{50 S_{1/2}}$ (dotted line). The intersection of grey lines fixes our chosen working point.}
\label{cfig:fig2}
\end{figure}
It is evident from Eqs.~\eqref{eq:ucprR}, \eqref{eq:modificationgamma} and \eqref{eq:atompolaziability} that, due to their large principal quantum numbers, the energy shift and relaxation rate are much larger for Rydberg atoms than in the case of regular atoms. As we show below, this fact is crucial for the achievement of resolved sidebands cooling protocol.

\textit{Coupling of Rydberg state and phonon modes.$-$}
Since the CP induced energy shifts are extremely sensitive to small distance fluctuations (it scales as $\sim (d+ \delta z)^{-3}$ in the nonretarded limit, with $\delta z$ denoting the departure from equilibrium), the out-of-plane motion of graphene introduces a modulation in the local transition frequencies. For small amplitudes, $ \delta z \ll d$, one can write the total Hamiltonian of the system as \cite{Muschik2014},
\begin{align}
\hat{H}= \hbar \omega_{m}\hat{b}^{\dagger}\hat{b}+\hbar \frac{\Omega}{2}\sigma_{x}
- \hbar \frac{\Delta}{2}\sigma_{z} + \hbar g(\hat{b}+\hat{b}^{\dagger})\sigma_{z}.
 \label{eq:Hamiltonian}
 \end{align}
The first term is the oscillator Hamiltonian, with $\hat{b} \pc{\hat{b}^\dagger}$ standing for the phonon annihilation (creation) operator; the second term describes the Rydberg laser driving of effective Rabi frequency $\Omega$ and detuning $\Delta$ with respect to Rydberg transition frequency $\omega_{a} = \omega_{0}+\Delta \omega (d)$ \cite{Muschik2014,bernien2017probing, Zhang2015}, and the surface-induced shift which can be calculated as $\Delta \omega (\delta z+d) = \delta\omega_{r}-\delta\omega_{g}$. Finally, the last term in Eq.~\eqref{eq:Hamiltonian} is the (electro-mechanical) interaction between the atom and the oscillating surface with effective coupling $g = g_{0} z_{\rm zpf}$. Here, $g_{0} =\partial_{z}\Delta\omega\vert_{z=d}$ is the CP coupling and $z_{\rm zpf} = \sqrt{\hbar/ 2 m \omega_{m}}$ is the zero-point fluctuation of the oscillator, characterized by its effective mass $m$ and frequency $\omega_{m}$; the position operator is given by $\hat{z}=z_{\rm zpf}(\hat{b}+\hat{b}^{\dagger})$. A similar situation involving spin-mechanical coupling has been investigated in a variety of setups, aiming at phonon control \cite{PhysRevA.46.2668, PhysRevLett.92.075507, PhysRevB.69.125339, 1367-2630-10-9-095019, PhysRevLett.102.096804, PhysRevLett.109.147205, PhysRevB.88.064105,  PhysRevLett.119.233602} or at efficient mechanical readout \cite{PhysRevLett.115.203601}. In the remainder of this manuscript, we show that the Hamiltonian in Eq.~\eqref{eq:Hamiltonian} allows for an unprecedented resolved sideband cooling of the graphene flexural mode. 

In Fig.~\ref{cfig:fig2}, we present the numerical results for the vacuum coupling $g_0$ and the surface-assisted decay rate $\Gamma$ as a function of the atom-surface distance $d$ for different Rydberg states $n$ of $^{87}$Rb. As one can immediately see, it is possible to achieve couplings of the same order of magnitude as in the case of a two-level emitter, where only a single transition $\vert g\rangle \leftrightarrow \vert e \rangle$ is taken into account for the calculations of energy shifts ($g_{0}\simeq 10^{11}$~Hz/nm for $d\simeq 20~\rm nm$ in Ref.~\cite{PhysRevLett.112.223601}). However, for the Rydberg case presented here, we incorporate all the allowed transitions $\textstyle\sum_{k}(\vert r\rangle \leftrightarrow \vert k \rangle)$ (with $\vert k \rangle$ denoting the states lying below and above the Rydberg state $\vert r \rangle$ as a consequence of the surface-induced degeneracy lifting - see Fig.~\ref{cfig:fig1}). As a result, equally large couplings are obtained for distances at least one order of magnitude larger. Moreover, while the free-space spontaneous decay rate of Rydberg state is of the order of a few kHz, the presence of a macroscopic surface significantly increase these values up to few MHz for such large distances due to the changes in the vacuum density of states. As we are about to see, similar to cavity-enhanced decay that assists cooling of polar molecules \cite{Andr2006}, these enhanced decay rates make sideband cooling feasible for graphene flexural modes.

The advantage of exploiting the Rydberg states for graphene cooling is then twofold: i) strong enough couplings achieved for sufficiently large atom-surface distances ($d\sim \mu$m) is a crucial element in real-life experiments, as it allows to place atoms near surfaces with standard quantum optics techniques without heating graphene \cite{PhysRevLett.103.123004, PhysRevA.97.063423, PhysRevLett.122.123602}; ii) the resulting surface-assisted decay $\Gamma$ falls below the mechanical frequency range ($\omega_{m}/2\pi\simeq 1-100 ~\rm MHz$), a central ingredient for the establishment of the resolved sideband condition. The latter is otherwise unavailable in similar protocols involving two-level emitters (see for instance Ref.~\cite{PhysRevLett.122.023602}). As such, we have tuned the atom-surface distance and chosen the Rydberg state number to obtain appropriate values of $g_0$ and $\Gamma$. In the remainder of our calculations, we have set $d=0.7 \mu \text{m}$ for the state $n=80$ of $^{87}$Rb, as depicted in Fig.~\ref{cfig:fig2}.

\begin{figure}[t!]
\includegraphics[width=\linewidth]{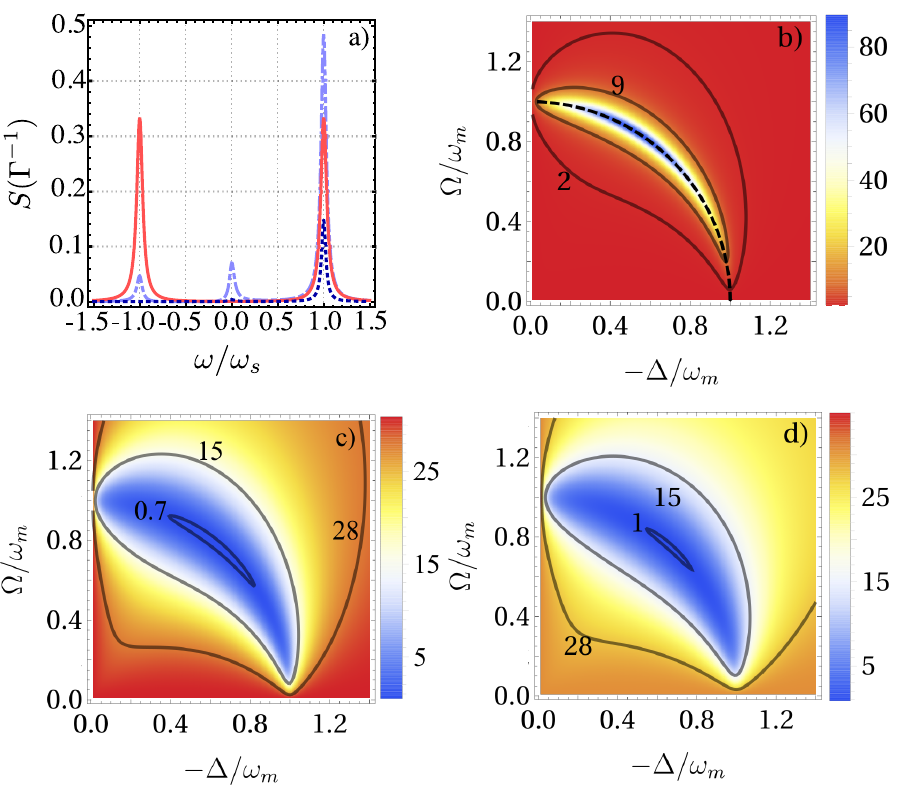}
\caption{(Color online) a) Real part of the atomic spectrum for $\Omega=0.85~\omega_{m}$ (dot-dashed line), $\Omega = \omega_m$ (solid line)  and $\Omega=0.4~\omega_{m}$ (dotted line), with peaks $\omega_s=\pm \sqrt{\Omega^2+\Delta^2}$. Optimal driving is achieved for $\Omega=0.85~\omega_{m}$. For $\Omega=\omega_{m}$, the heating and cooling peak balance each other, therefore, the coupled atomic system does not work as cold bath. b) Net cooling rate $\Gamma_{c}$ (in units of kHz) of the graphene resonator. c) Steady-state phonon number $n_{\rm ss}$. d) The same as in c) but considering a dephasing rate $\tilde{\Gamma}=0.015~\omega_{m}$. For b), c) and d), we have set $\omega_{m}= 2\pi \times 20$~MHz and $m=1.6\times10^{-18}$~kg.
\label{cfig:fig3}}
\end{figure}

\textit{Sideband cooling of the graphene flexural mode.$-$} 
Our cooling scheme is based on a reservoir engineering protocol, where phonons are scattered away thanks to a sufficiently fast relaxation of the atomic degrees of freedom (DOF), to which the graphene flexural mode is coupled with via the CP force. Similar reservoir-engineering schemes are at the basis of different ground-state cooling solutions, achieved for single ions \cite{PhysRevLett.62.403}, optomechanical setups \cite{PhysRevLett.99.093902, schliesser2008resolved, Teufel2011}, single atoms \cite{PhysRevLett.110.133001} and nitrogen-vacancy centres in diamond \cite{macquarrie2017cooling}. These have also been recently proposed for carbon nanotubes \cite{PhysRevLett.113.047201, PhysRevLett.102.096804, PhysRevB.95.205415}.

Taking advantage of the features of Rydberg atoms discussed above, we construct a resolved sideband cooling scheme for the fundamental flexural mode of a graphene resonator. We consider that the mechanical mode is initially in equilibrium with a (phonon) bath of temperature $T$, and thus the mean phonon number in the mode is given by $\bar{n}_{\rm th}=(\exp({\hbar \omega_{m} / k_{B}T})-1)^{-1}$. For our purpose, we will consider a graphene resonator of total mass $m=1.6\times 10^{-18}$~kg and quality factor $Q=2\times10^{5}$ \cite{weber2016force,chen2009performance, doi:10.1002/andp.201400153}. The graphene resonator of frequency $\omega_{m}= 2\pi \times 20$ MHz is kept at a cryogenic temperature $T=30$~mK (see e.g Ref.~\cite{weber2016force,doi:10.1002/andp.201400153}), for which we find $\bar{n}_{\rm th}\simeq 30$. The chosen values for the coupling $g_0=1.12$~GHz/nm (correspondingly, $g=0.57$~MHz) and decay $\Gamma= 3.24$~MHz allow us to operate within resolved sideband cooling regime, since the conditions $g/\Gamma\simeq 0.18 <1$ and $\Gamma/\omega_{m}\simeq 0.03<1$ are simultaneously satisfied, while keeping the strong cooperativity condition, $g^{2}/\Gamma (\gamma_{m} n_{\rm th})\simeq 5.2 >1$ \cite{1367-2630-10-9-095019, PhysRevLett.119.233602,macquarrie2017cooling, supp}. The dynamics of the open system is governed by Markov master equation 
\begin{equation}
\dot{\hat{\rho}} =-(i/\hbar) \pr{ \hat{H},\hat{\rho} } +\mathcal{L}(\hat{\rho}),
\end{equation}
where the Liouville operator accounting for irreversible dynamics due to the coupling with the various dissipative channels are given by
\begin{align}
\mathcal{L}(\hat{\rho})& =\frac{\gamma_{m}}{2}(\bar{n}_{\rm th}+1)\mathcal{D}_{\hat{b}}\pr{\hat{\rho}}+\frac{\gamma_{m}}{2}\bar{n}_{\rm th}\mathcal{D}_{\hat{b}^{\dagger}}\pr{\hat{\rho}}\nonumber
\\
&\quad +\Gamma\mathcal{D}_{\hat{\sigma}_{-}}\pr{\hat{\rho}}
+\frac{\tilde{\Gamma}}{4}\mathcal{D}_{\hat{\sigma}_{z}}\pr{\hat{\rho}},
\label{eq:mastereq} 
\end{align}
with $\mathcal{D}_{\hat{o}}\pr{\hat{\rho}}=2\hat{o}\hat{\rho}\hat{o}^{\dagger}-\hat{o}^{\dagger}\hat{o}\hat{\rho}-\hat{\rho}\hat{o}\hat{o}^{\dagger}$. The first and the second terms in Eq.~\eqref{eq:mastereq} describe the thermalisation process with the thermal bath, respectively accounting for the (stimulated and spontaneous) thermal emission and thermal excitation of the mechanical mode of vacuum decay rate $\gamma_{m}=\omega_{m}/Q$. The third one describes the atomic spontaneous emission in the presence of a graphene resonator $\Gamma$; the fourth term accounts for the dephasing occurring at the rate $\tilde{\Gamma}$. Note that for the cryogenic temperatures considered, we can safely neglect the atomic thermal excitation.

In the Lamb-Dicke regime, $g \sqrt{\bar{n}_{\rm th}+1}<\omega_{m}$ \cite{PhysRevLett.92.075507}, where the mode-atom interaction is only a perturbative effect,  the system will relax into the product state $ \hat{\rho}(t)\simeq\hat{\rho}_{\rm ss}\otimes\hat{\rho}_{m}(t)$, which will be composed of atomic steady-state density matrix $\hat{\rho}_{ss}$ and the reduced mechanical density matrix $\hat{\rho}_{m}=\Tr_{a}\pr{\rho}$, with the trace performed over the atomic degrees. This allows us to adiabatically eliminate the atomic degrees and re-write the master equation for the mechanical modes alone \cite{1367-2630-10-9-095019},
\begin{align}
\dot{\hat{\rho}}_{m}=-i \omega_{m}\pr{\hat{b}^{\dagger}\hat{b} ,\hat{\rho}_{m} }+A_{-}\mathcal{D}_{\hat{b}}\pr{\hat{\rho}_{m}}+A_{+}\mathcal{D}_{\hat{b}^{\dagger}}\pr{\hat{\rho}_{m}},
\label{eq:mastereqm}
\end{align}
where $A_{-}=g^{2} {S(\omega_{m})}+(\gamma_{m}/2)(\bar{n}_{\rm th}+1)$ and $A_{+}=g^{2} {S(-\omega_{m})}+(\gamma_{m}/2)\bar{n}_{\rm th}$ describe the effective flexural cooling and heating rates, respectively. These processes are a result of the coupling to the dissipative atomic bath, characterized by its steady-state spectral density \cite{breuer2002theory}
\begin{equation}
S(\omega)={\rm Re}\pr{\int_0^\infty d\tau e^{i\omega \tau}\langle\delta\sigma_{z}(\tau)\delta\sigma_{z}\rangle_{\rm ss}},
\end{equation}
evaluated at the mechanical frequency $\omega_m$. The spectrum is determined by the correlation of the atomic steady-state fluctuation operator $\delta\sigma_{z}=\sigma_{z}-\langle\sigma_{z}\rangle_{\rm ss}$ and obtained by solving Bloch's equations for the Rydberg atom alone (cf. Eq.~\ref{eq:mastereq}) and by applying the quantum regression theorem \cite{breuer2002theory}. The cooling and heating rates are evaluated at positive $\omega_{m}$ and negative $-\omega_{m}$ frequencies. This characterizes the ability of the atom to respectively absorb and emit phonons at that very frequency \cite{RevModPhys.82.1155}.
As one can see from Eq.~\eqref{eq:mastereqm}, one must enhance the phonon absorption in detriment of its emission, in order to increase the effective cooling rate. This means that we need to look at the parameters that enhance the positive frequency component of the (asymmetric) spectrum. In Fig.~\ref{cfig:fig3}~a), we have plotted the steady-state spectrum for different Rabi frequencies. As we can see, $S(\omega)$ exhibits three well resolved peaks at $\omega=0$ and $\omega=\pm\omega_{s}$. Here $\omega_{s}=\sqrt{\Omega^{2}+\Delta^{2}}$ is the energy splitting of the dressed atomic states $\ket{+}=\sin({\alpha/2})\ket{g}+\cos({\alpha/2})\ket{r}$ and $\ket{-}=\cos({\alpha/2})\ket{g}-\sin({\alpha/2})\ket{r}$, with $\tan{\alpha}=(\Omega/|\Delta|)$, as evaluated in the interaction picture (cf. Eq.~\ref{eq:Hamiltonian} and Ref.~\cite{Hauss2008,supp}). In the resolved sideband regime, $ \Gamma \ll \omega_{s}$, the optimal cooling conditions (i.e. when the phonon absorption capacity by the atom is maximized) is achieved at resonance $\omega_{s}=\omega_{m}$, for a driving strength of $\Omega\simeq 0.85~\omega_{m}$ \cite{PhysRevB.82.165320, 1367-2630-10-9-095019,supp}. In the interaction picture, the process $\ket{-,n}\mapsto \ket{+,n-1}$ corresponds to cooling, which is manifested by the spectrum peak at $\omega=\omega_{s}$. Such cooling is obtained for red detuned driving, $\omega_l-\omega_a=-\Delta$. From Eq.~\eqref{eq:mastereqm}, we obtain the mean flexural number,
\begin{equation}
\langle \dot{n} \rangle\equiv \Tr \pr{\dot{\hat{\rho}}_{m}\hat{b}^{\dagger}\hat{b}}=-\Gamma_{c}\langle n\rangle+A_{+},
\label{eq_rate}
\end{equation}
where the net cooling rate is given by $\Gamma_{c}=A_{-}-A_{+}$. Ground-state cooling is achieved if the steady-state flexuron number, $n_{\rm ss}=A_{+}/ \Gamma_{c}$, satisfies the condition $n_{\rm ss} < 1$. Using Eq.~\ref{eq:mastereqm} leads us the optical cooling rate $\gamma_{\text{opt}}$ and the minimal occupation number $n_{0}$ (see Ref.~\cite{supp}), reading
\begin{align}
\gamma_{\text{opt}} = 2g^{2}[S(\omega_{m})-S(-\omega_{m})], ~n_{0}=\frac{S(-\omega_{m})}{S(\omega_{m})-S(-\omega_{m})}.
\label{}
\end{align}
The net cooling rate $\Gamma_{c}$ and steady-state phonon number $n_{\text{ss}}$ then take the form,
\begin{align} 
\label{eq:netopticalcooling}
\Gamma_{c}=\frac{1}{2}\left(\gamma_{\text{opt}}+\gamma_{m}\right) , \quad  n_{\text{ss}}=\frac{\gamma_{\text{opt}}n_{0}+\gamma_{m}\bar{n}_{\rm th}}{\gamma_{\text{opt}}+\gamma_{m}}.
\end{align}
Since $\gamma_{\text{opt}}\sim(g^{2}/\Gamma)\gg \gamma_{m}$ \cite{supp}, one can write $n_{\text{ss}}\simeq n_{0}+(\gamma_{m}\bar{n}_{\rm th}/{\gamma_{\text{opt}}})$. Given that the heating process (related to the peak $S(-\omega_{m})$) is highly suppressed, the final occupation number is limited by the smallness of the ratio $\gamma_{m}\bar{n}_{\rm th}/{\gamma_{\text{opt}}}$, which is guaranteed for large values of $g$. The high polarizability of the Rydberg state provides such a requirement. In Fig.~\ref{cfig:fig3}~b), we plot $\Gamma_c$, displaying a maximum for the parameters contained by the dotted quarter circle. Moreover, the steady-state phonon number is shown in Fig.~\ref{cfig:fig3}~c). For the optimal cooling conditions, we get a record value $n_{\rm ss}\lesssim 0.7$. 

Finally, we investigate how robust the ground-state cooling protocol can be in the presence of dephasing, which may enter the cooling dynamics as a result of the atomic trap fluctuations and surface-induced noise. In fact, provided that the condition $\tilde{\Gamma}<\omega_{m}$ is satisfied, ground-state cooling is still attained. This fact is patent in Fig.~\ref{cfig:fig3}~d), revealing that $n_{\rm ss}\lesssim 1$ for $\tilde{\Gamma}=0.015~\omega_{m}$,  which is estimated to be met in typical conditions \cite{Li2013, PhysRevA.82.052517}. These dephasing effects will set a cooling threshold, clearly shortening the region of parameters for which ground-state cooling is achievable. This is therefore desirable to understand at which extent dephasing effects can be mitigated or controlled in state-of-the-art experimental conditions, although our calculations seem to suggest that our protocol is quite robust. 

\textit{Conclusions.$-$} We have constructed a protocol enabling ground-state sideband cooling of out-of-plane (flexural) mode in suspended graphene based on vacuum forces. Our setup consists of a driven Rydberg atom placed a few micrometres away from a graphene nanoresonator, which is coupled with the help of CP forces. Given the high-atomic polarizability of Rydberg atoms, the conditions for resolved sideband cooling are achievable, enabling to bring the fundamental flexural mode down to the quantum limit. Our findings overcome the difficulty associated to cooling schemes based on optomechanical laser cooling (a fact inherited from the large light absorbency of graphene in usual optomechanical interfaces), thus paving the way towards practical applications in quantum technology for which quantum motion is necessary. Moreover, this may also contribute to the enrichment of condensed matter platforms, as flexural modes are known to strongly affect the electronic mobility is suspended graphene \cite{PhysRevLett.105.266601}. For example, we anticipate that our scheme could be used to control the emission of graphene plasmons \cite{tielrooij2015electrical}, eventually potentiating the generation of plasmon lasers and control for the emission properties of atomic Ryberg states \cite{chang2017constructing, Yang2018}.

\textit{Acknowledgments.$-$} MK and HT acknowledge support from Funda\c{c}\~ao para a Ci\^encia e Tecnologia (FCT-Portugal) through Grant No PD/BD/114345/2016 and through Contract No IF/00433/2015 respectively. SR thanks the EPSRC grant EP/R002061/1.

\bibliographystyle{apsrev4-1}
\bibliography{References.bib}


\widetext
\newpage
\begin{center}
\textbf{\large Supplemental Material: Sideband ground-state cooling of graphene with Rydberg atoms via vacuum forces}
\end{center}
\setcounter{equation}{0}
\setcounter{figure}{0}
\setcounter{table}{0}
\setcounter{page}{1}
\makeatletter
\renewcommand{\theequation}{S\arabic{equation}}
\renewcommand{\thefigure}{S\arabic{figure}}
\renewcommand{\bibnumfmt}[1]{[S#1]}
\renewcommand{\citenumfont}[1]{S#1}

\subsection{Resonant and non-resonant part contribution in the energy shifts}

We briefly state about the origin of resonant and non-resonant contributions to the energy shifts described in the main text. The presence of a nearby surface changes the electromagnetic environment of the atom and results in substantial energy shifts (in comparison with the Lamb shift) and transition broadenings. The distance-dependent energy shift (Casimir-Polder shift) in the ground-state, $\delta\omega_{g}$, is due to non-resonant terms ($\hat{\sigma}_{gk} \hat{a}_{l}^{\dagger}$ and $\hat{\sigma}_{kg} \hat{a}_{l}$) of the interaction Hamiltonian $H_{\rm int}=-\mathbf{d}\cdot\mathbf{E}$. Here, $\hat{a}_{l}$ denotes the annihilation operator of the scattered electromagnetic field $\mathbf{E}$. In addition, the atomic operator $\sigma_{gk}=\ket{g}\bra{k}$ describes the transitions from the ground-state $\ket{g}$ to the intermediate states $\ket{s}$ (see Fig.~\ref{cfig:fig1}). In the single-photon Hilbert space, $H_{\rm int}$ couples the polaritonic ground ($\ket{g,0_{l}}$) and the intermediate ($\ket{k, 1_{l}}$) states. These excitations are followed by the re-absorption process of a photon by the atom, bringing the atom back to its ground-state (the field mode to its vacuum state). As these processes violate energy conservation, they are purely virtual ensuring that the ground-state energy shift is due to non-resonant part of the interaction as stated in the main text. 

For the Rydberg state, however, the transition manifold contains both above- and low-lying states, denoted by $\vert k \rangle$ in Fig.~\ref{cfig:fig1}. As a result, the corresponding shift $\delta\omega_{r}$ contains both non-resonant and resonant terms (thus coupling the polaritonic $\ket{r,0_{l}}$  and $\ket{k,1_{l}} $ states). While former accounts for virtual processes, the latter allow real photon transitions. The total shift $\delta\omega_{r}$ is then the sum of non-resonant ($\delta\omega_{r}^{\rm NR}$) and resonant ($\delta\omega_{r}^{\rm R}$) parts and the expression are given in main text.
 
\subsection{Surface-scattered Green's function}

The scattered part the electromagnetic field is characterized by the dyadic Green's function $ \mathbf{G}$, which is the solution of Helmholtz equation $[(\mathbf{\nabla} \times\mathbf{\nabla} \times )-(\omega^2/c^2)\epsilon(\textbf{r},\omega)]\mathbf{G}(\textbf{r},\textbf{r}',\omega)=\delta(\textbf{r}-\textbf{r}')\otimes\mathbb{I}$, solved for the planar geometry. Evaluating it at a transverse distance $z$, it encompass the effects of the graphene surface via its optical conductivity $\sigma$ through the transverse electric $R_{\mathrm{\rm TE}}$  ($s$-polarized)  and transverse magnetic $R_{\mathrm{\rm TM}}$ (p-polarized) reflection coefficients \cite{Dispersion Forces II},
\begin{equation}
\mathbf{G} \pc{z, z, \omega} = \frac{i}{8 \pi} \int_{0}^{\infty} d k_{\parallel} \frac{k_{\parallel}}{k_{\perp}} e^{ 2i k_{\perp} z}\text{Diag} \pr{R_{\mathrm{TE}}-\left(\frac{c^{2}k_{\perp}^{2}}{\omega^{2}}\right) R_{\mathrm{TM}}, R_{\mathrm{TE}}-\left(\frac{c^{2}k_{\perp}^{2}}{\omega^{2}}\right) R_{\mathrm{TM}},\left(\frac{2c^{2}k_{\parallel}^{2}}{\omega^{2}}\right)R_{\mathrm{TM}}}.
\label{eq_green}
\end{equation}
Here, $\rm{Diag}\pr{.,.,.}$ is a $3\times3$ diagonal matrix. The free-space wavevector satisfies $k^2=k_{\parallel}^{2}+k_{\perp}^{2}=(\omega^{2}/c^{2})$, while $k_{\perp}'=\sqrt {\epsilon k^{2}-k_{\parallel}^{2}}$ would be the wavevector in the graphene plane (for suspended graphene, $\epsilon=1$). The Fresnel refection coefficients  $R_{\mathrm{ TE}}=R_{\mathrm{\rm TE}}(\omega, k_{\parallel})$,  $R_{\mathrm{ TM}}= R_{\mathrm{TM}}(\omega, k_{\parallel})$  can be found by applying appropriate boundary conditions, yielding \cite{D. E. Chang},
\begin{align}
R_{\mathrm{TM}}=\frac{\epsilon k_{\perp}-k_{\perp}'+4 \pi \sigma k_{\perp}k_{\perp}'/\omega }{\epsilon k_{\perp}+k_{\perp}'+4 \pi \sigma k_{\perp} k_{\perp}'/\omega}, \quad R_{\mathrm{TE}}=\frac{ k_{\perp}-k_{\perp}'+4 \pi \sigma k/c}{ k_{\perp}+k_{\perp}'+4 \pi \sigma k/c}.
\end{align}
To calculate the non-resonant part of the energy shift, we have to introduced the imaginary frequency $\omega = i \xi$ allowing for a better evaluation of the poles \cite{Dispersion Forces II}). For the real-frequency dependence, the two regions $0<k_{\parallel}<\omega/c$  (propagating waves) and $k_{\parallel}>\omega/c$ (evanescent waves) are considered in the numerical computation of \eqref{eq_green}.  
\begin{figure*}[!t]
 \includegraphics[width=0.42\linewidth]{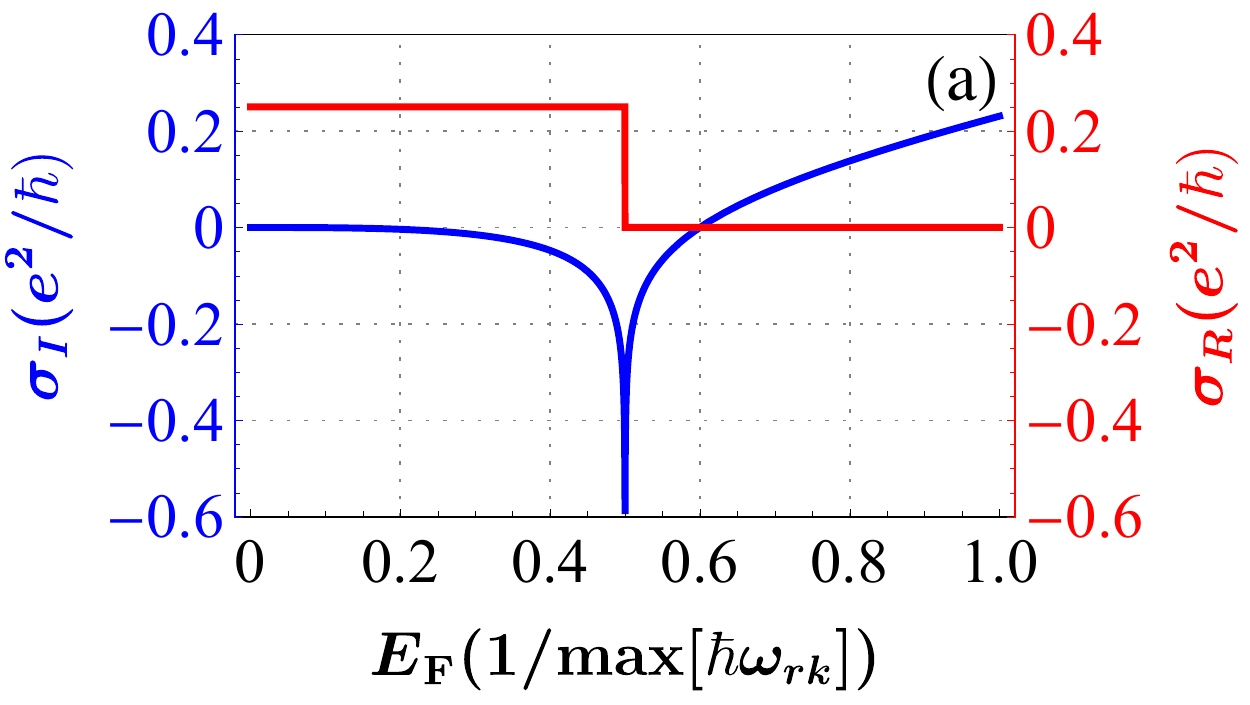} 
 \hspace*{1cm}
 \includegraphics[width=0.35\linewidth]{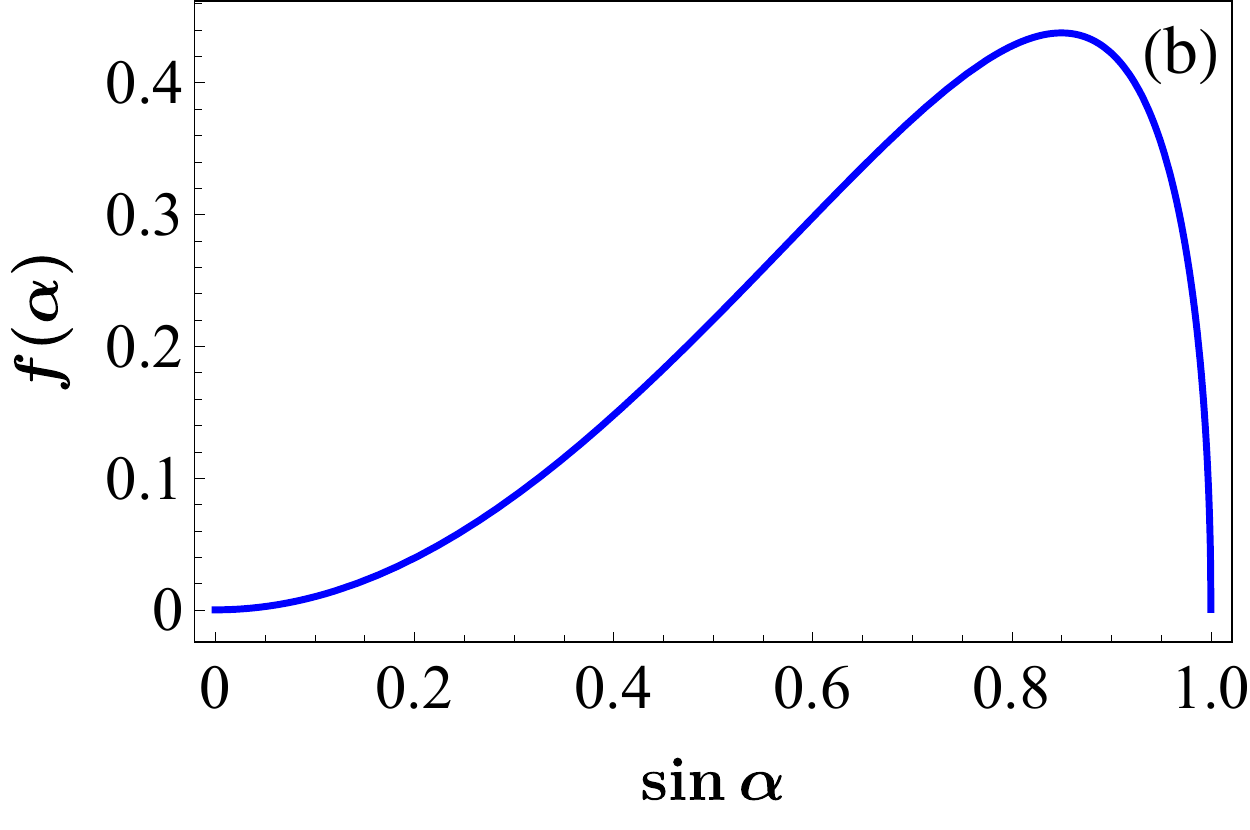}
 \caption{(Color online) (a) Photonic conductivity behaviour of graphene. (b) Dimensionless function $f (\alpha) $.}\label{scfig:fig1}
\end{figure*}

\subsection{Graphene properties}

The graphene conductivity $\sigma=\sigma(\omega)$ is given within the Random Phase Approximation (RPA), whose real $\sigma_{R}(\omega)$ and imaginary $\sigma_{I}(\omega)$ parts takes the form \cite{D. E. Chang}, 
\begin{equation}
\sigma_{R}(\omega)=\frac{e^{2}E_{\text{F}}}{\pi\hbar^{2}}\frac{1/\tau}{\omega^{2}+1/\tau^{2}}+\frac{e^{2}}{4\hbar}\theta(\hbar\omega-2E_{\text{F}}), 
\end{equation}
\begin{equation}
\sigma_{I}(\omega)=\frac{e^{2}E_{\text{F}}}{\pi\hbar^{2}}\frac{\omega}{\omega^{2}+1/\tau^{2}}-\frac{e^{2}}{4\pi\hbar}\text{log}\left|\frac{2E_{\text{F}}+\hbar\omega}{2E_{\text{F}}-\hbar\omega}\right|,
\end{equation}
where $\tau=1/\gamma_{g}$ denotes the electron scattering time. The typical behavior of both the real and imaginary parts of the optical conductivity is shown in Fig.~\ref{scfig:fig1}~a). For $E_{\rm F}\lesssim 0.5~ \text{max}[\hbar \omega_{rk}]$, the conductivity is mostly real and the atomic transition decay dumps energy into the particle-hole (inter-band) continuum. In the region, $0.5 ~\text{max}[\hbar \omega_{rk}]\lesssim E_{\text{F}}\lesssim ~0.7~\text{max}[\hbar \omega_{rk}] $, the atom decays into the free space, while for $E_{\text{F}}\gtrsim  0.7~\text{max} [\hbar \omega_{rk}]$ the decay process results in a plasmon excitation (intra-band), as witnessed by a positive imaginary part in the conductivity \cite{L. Orona}.

It is well known, and also verified by our numerics, that the shift in the state $n$ is mainly due to the transition $|nS_{1/2}\rangle \mapsto |nP_{1/2,3/2}\rangle $, i.e. the transition of lowest frequency (i.e. $\text{min} [\hbar \omega_{rk}]\simeq0.031~\text{meV}$) and largest dipole moment \cite{S.A.Ellingsen}. For $n=80$, the choice of the Fermi energy in the main text ($E_{\text{F}}=0.8~\text{eV}$) suggests that emissions in these frequency range result in plasmon excitations for almost $\simeq 97\%$ of whole the transition manifold, as the condition $E_{\text{F}}\gtrsim  0.7[\hbar \omega_{rk}]$ is fulfilled. Therefore, only a small percentage of the transitions (and thus with a negligible contribution to the atomic energy shifts) will emit in the particle-hole continuum. Both intra- and inter-band emissions could potentially heat graphene, via the sequential excitation of in-plane phonons. However, their effect in the heating via the production of out-of-plane phonons (flexurons) are unknown. Although interesting, the latter are out of the scope of present work.

%
\subsection{Atomic dressed states}
We start by considering the Hamiltonian appearing in Eq.~(6) of the manuscript ($\hbar=1$),
\begin{align}
\hat{H}= \omega_{m}\hat{b}^{\dagger}\hat{b}+ \frac{\Omega}{2}\sigma_{x}
-  \frac{\Delta}{2}\sigma_{z} + g(\hat{b}+\hat{b}^{\dagger})\sigma_{z},
 \end{align}
and transforming it into the basis in which the two-level atom is diagonal (Dicke basis),
 \begin{align}
\hat{H}= 
- \frac{\omega_{s}}{2}\sigma_{z} + \omega_{m}\hat{b}^{\dagger}\hat{b}+ [g\cos(\alpha) \sigma_{z}-g\sin({\alpha})\sigma_{x}]  (\hat{b}+\hat{b}^{\dagger}),
\label{eq_INT}
 \end{align}
 where $\omega_{s}=\sqrt{\Delta^2+\Omega^{2}}$ is the energy splitting of two dressed state $\ket{-}, \ket{+}$ discussed in the text. Here, $\sin{(\alpha)}=\Omega/\omega_{s}$ and $ \cos{(\alpha)}=-\Delta/\omega_{s}$. Then, we move to the interaction picture via the unitary transformation $\hat U=\exp[i(\omega_{m}\hat{b}^{\dagger}\hat{b}+(\omega_{s}/2)\sigma_{z})t]$. After performing a rotating wave approximation (RWA) (valid for $g\ll\omega_{s}=\omega_{m}$), we recast Eq.~\eqref{eq_INT} as
\begin{align}
\hat{H}_I=\hat U^\dagger \hat H \hat U =
- \tilde{g} (\sigma_{-}\hat{b}^{\dagger}+\sigma_{+}\hat{b}),
\label{eq_RWA}
 \end{align}
with $\tilde{g}=g\sin({\alpha})$. In the single-phonon sector, the Hamiltonian in Eq.~\eqref{eq_RWA} drives the transition $\ket{-,n} \leftrightarrow\ket{+,n-1}$. Since the atomic decays are fast, $\Gamma>\tilde g$, the latter is over damped, which guarantees that phonons can be damped away thanks to the dressing with the atomic degrees of freedom. 

\subsection{Atomic power spectrum}

Coming back to Eq.~(7) of main text, we can obtain dynamics of Bloch vector $\vec{\sigma}=\pr{\sigma_{x},\sigma_{y},\sigma_{z}}^T$ from the isolated Bloch equation, i.e. considering only the atomic part of the Hamiltonian, thus taking $\gamma_{m}=0$. In the limit of pure atomic decay (i.e. $\tilde{\Gamma}\mapsto 0$), the Bloch vector evolves as
\begin{align} 
\partial_{t}\langle\vec{\sigma}\rangle = A\langle\vec{\sigma}\rangle-\vec{\Gamma},
 \end{align}
where 
\begin{equation}
A=
\left(
\begin{array}{cccc}
{-\Gamma}&{-\omega_{s}\cos(\alpha)}&{0}\\
{\omega_{s}\cos(\alpha)}&{-\Gamma}&{-\omega_{s}\sin(\alpha)}\\
{0}&{\omega_{s}\sin(\alpha)}&{2\Gamma}\\
\end{array}
\right), \quad
\vec{\Gamma}=
\left(
\begin{array}{c}
{0}\\
{0}\\
{2\Gamma}\\
\end{array}
\right).
\end{equation}
Using the quantum regression theorem \cite{K. Hammerer}, one can write the spectrum $\mathcal{S}(\omega)=\langle\delta\sigma_{z}(-i\omega)\delta\sigma_{z}(0)\rangle_{\rm ss}$ for the steady-state correlation of the fluctuation operator $\delta\sigma_{z}$ (defined in the main text) as
\begin{equation}
\mathcal{S}(\omega)=-(0,0,1)\cdot \pr{i\omega \textbf{1}+A}^{-1}\vec{B},
\end{equation}
where $\vec{B}=\langle\delta\vec{\sigma}\delta\sigma_{z}\rangle_{\rm ss}$. Given the inverse of the matrix $\pr{i\omega \textbf{1}+A}^{-1}=\text{Adj}\pr{i\omega \textbf{1}+A}/\text{Det}\pr{i\omega \textbf{1}+A}$ (with $\text{Adj} \pr{.}$ and $\text{Det}\pr{.}$ denoting the adjugate and determinant, respectively), one can write
\begin{equation}
\mathcal{S}(\omega)=\frac{h(\omega)}{(i\omega+\epsilon_{0})( i\omega+\epsilon_{+})(i\omega+\epsilon_{-})},\quad \text{where}\quad h(\omega)=-(0,0,1)\cdot \text{Adj}\pr{i\omega \textbf{1}+A}^{-1}\vec{B}.
\end{equation}
Here, $\epsilon_{i}$ are the eigenvalues of matrix $A$. The spectrum generally exhibits three peaks, whose frequencies and width are respectively given by the imaginary and real parts of $\epsilon_i$. In the resolved sideband regime, $\Gamma/\omega_{s}\ll 1$, the eigenvalues values are given by
\begin{equation}
\epsilon_{0}\simeq -\Gamma_{\epsilon_0},\quad  \epsilon_{\pm}\simeq \pm i\omega_{s}- \Gamma_{\epsilon_+},
\end{equation}
where,
\begin{equation}
\Gamma_{\epsilon_0}=\frac{ \Gamma}{2}(\cos (2 \alpha )+3), \quad \Gamma_{\epsilon_+}=\frac{\Gamma}{4}(\cos (2 \alpha )-5).
\end{equation}
It is therefore clear that the spectrum exhibits well-resolved peaks at $\omega=(0,\pm\omega_{s})$, as illustrated in Fig.~3 of the manuscript. In the same regime, the real part of the spectrum in the vicinity of $\pm \omega_s$ reads,
\begin{equation}
 S(\omega)\equiv\text{Re} \pr{{\mathcal{S}(\omega\simeq \pm\omega_s)}}=\frac{\Gamma_{\epsilon_+}}{\Gamma_{\epsilon_+}^{2}+(\omega\mp\omega_{s})}(P_{\pm}), \quad \text{with} \quad P_{\pm}=\text{Re} \pr{\frac{-h(\pm\omega_{s})}{(\pm i\omega_{s}-\Gamma_{\epsilon_0})( \pm 2i\omega_{s}-\Gamma_+)}},
\end{equation}
To the lowest order in $\Gamma/\omega_{s}$, $P_{\pm}$ is explicitly given by
\begin{equation}
 P_{+}=\frac{\sin ^6(\alpha ) \csc ^4\left(\frac{\alpha }{2}\right)}{4 (\cos (2 \alpha )+3)},  \quad P_{-}=\frac{4 \sin ^4\left(\frac{\alpha }{2}\right) \sin ^2(\alpha )}{\cos (2 \alpha )+3}. 
\end{equation}
For convenience, we  finally define a dimensionless function $ f(\alpha)=\Gamma [S(\omega)-S(-\omega)]$ capturing the dependence of the net cooling rate on the relevant parameters of the problem. In the limit $\omega\rightarrow \omega_{s}$, the function turns out to be
\begin{equation}\label{eq:spectrumapprox}
f(\alpha) =-\frac{8 \sin (\alpha ) \sin (2 \alpha )}{ (\cos (2 \alpha )-5) (\cos (2 \alpha )+3)}. 
\end{equation}
As it can be seen in Fig.~\ref{scfig:fig1}, the latter is peaked at $\Omega\simeq 0.85 \omega_m$, revealing the optimal configuration for resolved sideband cooling to take place. 

\subsection{Contribution by coupled dissipative atomic bath and net cooling rate}

From Eq.~(8) of the main text, one can isolate the effects of the thermal bath (the uncontrolled reservoir) and that of the ``dissipative atomic bath" (controlled reservoir). This allows us to write the equation in the form
\begin{align}
\dot{\hat{\rho}}_{m}=-i \omega_{m}\pr{\hat{b}^{\dagger}\hat{b} ,\hat{\rho}_{m} }+	\frac{\gamma_{\text{opt}}}{2}(n_{0}+1)\mathcal{D}_{\hat{b}}\pr{\hat{\rho}_{m}}+\frac{\gamma_{\text{opt}}}{2} n_{0}\mathcal{D}_{\hat{b}^{\dagger}}\pr{\hat{\rho}_{m}}+ \	\frac{\gamma_{m}}{2}(\bar{n}_{\rm th}+1)\mathcal{D}_{\hat{b}}\pr{\hat{\rho}}+\frac{\gamma_{m}}{2}\bar{n}_{\rm th}\mathcal{D}_{\hat{b}^{\dagger}}\pr{\hat{\rho}},
\end{align}
where $\gamma_{\text{opt}}=2g^{2}[S(\omega_{m})-S(-\omega_{m})]$ is the optical cooling rate and $n_{0}=S(-\omega_{m})/[S(\omega_{m})-S(-\omega_{m})]$ is the minimal (or residual) flexural occupation number. Both quantities are a result of the coupling to the dissipative atomic bath alone. Conversely, the net cooling rate $\Gamma_{c}$ and the steady-state phonon number $n_{\text{\rm ss}}$ described in the manuscript take the form
\begin{align} \label{eq:netopticalcooling}
\Gamma_{c}=	\frac{1}{2}\left(\gamma_{\text{opt}}+\gamma_{m}\right), \quad  n_{\text{ss}}=\frac{\gamma_{\text{opt}}n_{0}+\gamma_{m}\bar{n}_{\rm th}}{\gamma_{\text{opt}}+\gamma_{m}}.
\end{align}
Using Eq.~\ref{eq:spectrumapprox}, and taking $\omega_{s}=\omega_{m}$, the total optical cooling rate reads,
\begin{align}\label{eq:netopticalcoolingapprox}
\gamma_{\text{opt}}\simeq (2g^{2}/\Gamma) f(\alpha)=\frac{8\sin^{2}(\alpha)\sqrt{1-\sin^{2}(\alpha)}}{4-\sin^{4}(\alpha)}\frac{g^{2}}{\Gamma}.
\end{align}
The net optical cooling rate turns out to be on the order of $\mathcal{O} (g^{2}/\Gamma)$. To achieve ground-state cooling, $\gamma_{\text{opt}}$ must exceed the flexural thermal heating  rate $\gamma_{m}\bar{n}_{\rm th}$. As a consequence, a large value of the ratio $(\gamma_{\text{opt}}/\gamma_{m}\bar{n}_{\rm th})=g^{2}/\Gamma\gamma_{m}\bar{n}_{\rm th}$ (dubbed the cooperativity parameter in the literature) is required to achieve ground-state cooling \cite{Rabl2010}.

\bibliographystyle{apsrev4-1}

\end{document}